\documentstyle[graphics,psfig]{aa}

\begin{document}

\title{XMM-Newton study of the persistent X-ray source 1E~1743.1--2843 
located in the Galactic Center direction}
\author{D. Porquet\inst{1} \and J. Rodriguez\inst{1,2} \and S. Corbel\inst{1,3} 
\and P. Goldoni\inst{1} \and R.S. Warwick\inst{4} \and A. Goldwurm\inst{1} 
\and A. Decourchelle\inst{1} } 

\offprints{Delphine Porquet,\\
\email{dporquet@mpe.mpg.de (current address)}}

\institute{Service d'Astrophysique, CNRS URA 2052, CEA Saclay, F-91191 Gif-sur-Yvette, France
\and Integral Science Data Center, Chemin d'Ecogia, 16 CH-1290 Versoix, Switzerland
\and Universit\'e Paris VII, f\'ed\'eration APC, 2 place Jussieu, F-75005 Paris Cedex, France 
\and X-Ray Astronomy Group; Department of Physics and Astronomy; Leicester 
University; LE1 7RH; U.K.}
\date{Received ...  / accepted ...}

\abstract{ We  report the results of an {\sl XMM-Newton} observation of
the persistent X--ray  source \object{1E~1743.1--2843}, located in the
Galactic  Center (GC)  direction.  
 We determine the position of the source at 
$\alpha_{\rm J2000}$=17$^{h}$\,46$^{m}$\,21.0$^{s}$, 
$\delta_{\rm J2000}$=$-$28$^{\circ}$\,43$^{\prime}$\,44$^{\prime\prime}$ 
 (with an uncertainty of 1.5$^{\prime\prime}$), which is the most accurate to date, 
 and will enable cross-identifications  at other wavelengths.
The  source was  bright  during this
observation  (L$_{\rm  2-10\,keV}$  $\sim$  2.7  $\times$  10$^{36}$
  d$^{2}_{\rm 10kpc}$ erg\,s$^{-1}$ for a power-law continuum), 
  with no significant variability.  
We propose  that 1E~1743.1--2843 may be
explained in terms of  a black hole  candidate in  a low/hard state.   
There is  an indication  that the  source  exhibits different
states  from a comparison  of our  results with  previous observations
(e.g.,  {\sl ART-P},  {\sl BeppoSAX}).  
However, the present spectral analysis does not rule out the
 hypothesis of  a neutron star low-mass  X-ray binary as
suggested previously.  
If 1E~1743.1--2843 is actually  located in the GC region, we
might  expect to  observe significant   6.4\,keV  fluorescent iron
line emission from nearby molecular clouds (e.g., GCM+0.25+0.01).
   }   

\titlerunning{XMM-Newton   study   of   1E~1743.1--2843}
\authorrunning{Porquet et al.}  \maketitle

\section{Introduction}

The  source 1E~1743.1--2843  was  discovered during  the first  X--ray
imaging observations of the Galactic Center (GC) region performed with
the {\sl  Einstein} Observatory (Watson et  al. \cite{Watson81}).  Its
column density ($> 10^{23}$ cm$^{-2}$)  is one of the highest observed
in the bright X--ray sources found in this region of sky, suggesting a
distance  similar  to,  or  greater, than  the  GC  
 (d=7.9$\pm$0.3 kpc, McNamara et al. \cite{McNamara2000}).
1E~1743.1--2843 has  been detected by all X-ray  satellites with X-ray
imaging capability above 2  keV (Watson et al.  \cite{Watson81}, Kawai
et al.  \cite{Kawai88}, Sunyaev et al.  \cite{Sunyaev91}, Pavlinsky et
al.   \cite{Pavlinsky94},   Lu  et  al.    \cite{Lu96},  Cremonesi  et
al. \cite{Cremonesi99}), whereas in  soft X-rays  (e.g. {\sl ROSAT}) the
source  is not detectable  due to  the high  column density  along the
line-of-sight (see Predehl \& Tr{\"u}mper \cite{Predehl94}).  Kawai et
al.  (\cite{Kawai88})  suggested that part of  the measured absorption
could  be  intrinsic,  as  is  the  case  for  \object{Vela  X-1}  and
\object{GX~301-2}.   The  inferred  X-ray 2--10\,keV luminosity of  this  bright
persistent  source  is about 2$\times$10$^{36}$$\times$d$_{10\rm{kpc}}^2$  erg\,s$^{-1}$
 ruling out models involving coronal or wind emission from
normal  stars but, conversely,  strongly favoring  the presence  of an
accreting  compact  object.   Cremonesi et  al.   (\cite{Cremonesi99})
suggested that  the absence  of periodic pulsations  (and/or eclipses)
and the relatively  soft X-ray spectrum favor a  low mass X-ray binary
(LMXB)  containing a  neutron  star.   LMXBs are  systems  in which  a
compact object (either a neutron star or a black hole) accretes matter
from  a  low-mass  ($<$  1\,M$_{\odot}$) companion  star.   Most  LMXBs 
containing neutron stars are characterized by the occurrence of Type~I
X--ray bursts  produced by the  thermonuclear flashes of  the accreted
material on the surface of  the neutron star.  However, no bursts have
been observed  from 1E~1743.1--2843 in extensive  observations over the
last 20 years.  If 1E~1743.1--2843 is a neutron star LMXB, the lack of
bursts     is    noteworthy     because    its     X-ray    luminosity
(10$^{36}$--10$^{37}$ erg  s$^{-1}$) is in the range  which is typical
of X--ray  bursters.  The  lack of bursting  activity in  neutron star
LMXB of higher luminosity is  generally ascribed to the stable burning
of  H  and  He in  sources  operating  close  to the  Eddington  limit
(Fujimoto et  al.  \cite{Fujimoto81}).   Such a high  luminosity would
require 1E~1743.1--2843 to be at  a distance greater than several tens
of kiloparsecs.  Type-I bursts are also suppressed in pulsars
due  to  the  higher   surface  magnetic   fields  (e.g.,   Lewin  et
al. \cite{Lewin95}).  Cremonesi al. (\cite{Cremonesi99})  did not rule
out  other  interpretations  such  as an  extra-galactic  source  seen
through  the Galactic  plane.\\ 

Here  we present the results of  the first observation 
of 1E~1743.1--2843 with {\sl XMM-Newton}. 
 Section 2 details the observation and data 
 reduction procedures. Section 3 presents the determination 
 of the accurate X-ray position of this object. 
Sections 4 and 5 describe, respectively, the timing and spectral 
 analysis. The analysis results are discussed 
in the last section.  

\section{Observations and data analysis}

1E~1743.1--2843 was observed by  {\sl XMM-Newton} on September 19, 2000
about  5.5$^{\prime}$ off-axis from  the center  of the  pointing. The
EPIC-MOS cameras were operated in the standard full-frame mode 
 (time resolution: 2.6\,s), 
 and the EPIC-PN camera in the extended full frame mode (time resolution: 200\,ms),
 with the medium filter used   in  both   cases.   
The  effective   exposure  times   were
$\sim$29.1\,ksec  and  $\sim$22\,ksec  for  the MOS  and  PN  cameras,
respectively.  The  data were reprocessed  using version 5.3.3  of the
Science  Analysis  Software  (SAS)  and further  filtered  using  {\sc
xmmselect}.  The  datasets were screened by rejecting  periods of high
background arising from marked increases  in the incident flux of soft
protons.  After this  data cleaning,  the useful  observing  times are
respectively  for MOS1  and MOS2  about 22.2\,ksec  and  23\,ksec, and
18.4\,ksec for PN.  Unfortunately, in the MOS1 CCD, 1E~1743.1--2843 
  is  located on  a  bright pixel  column  which  makes the  data
difficult to process.  Our present analysis is based largely on the PN
 data for  which we use single events  (corresponding to pattern 0)
to avoid  any possibility of  pile-up effects, although the  MOS2 data
(event  patterns   0--12)  did  provide   a  valuable  check   of  the
results.  The inferred  PN flux  in the  2--10\,keV  energy range is then 
about  2$\times$10$^{-10}$\,erg\,cm$^{-2}$s$^{-1}$,  thus  implying  a
negligible  pile-up fraction  ($<$2\%, see  Fig.~98 in  the XMM-Newton
Users' Handbook\footnote{{\tiny
http://xmm.vilspa.esa.es/external/xmm\_user\_support/documentation/uhb/}}
).\\
Counts from 1E~1743.1--2843 were extracted  within a radius of 
26.4$^{\prime\prime}$, thereby avoiding a gap between the PN CCDs.  
We  extracted   a  source-free   local  background
from an  annulus  centered  on  1E~1743.1--2843 with inner and outer radii
of 2$^{\prime}$ and 5$^{\prime}$ respectively.

\section{X-ray position and multi-wavelength counterpart}

In the field of view of our pointing, a bright X-ray point source 
 is associated to an optical foreground Tycho-2 source (\object{HD 316297}). 
 This very accurate position allow to  refine the astrometry and we find 
 for 1E\,1743.1--2843 the following position: 
$\alpha_{\rm J2000}$=17h\,46m\,21.0s, 
$\delta_{\rm J2000}$=$-$28$^{\circ}$\,43$^{\prime}$\,44$^{\prime\prime}$,   
 with a final accuracy  limited by
 the systematic residual uncertainties of 1.5$^{\prime\prime}$ (Kirsch \cite{Kirsch2002}).
  The  position inferred  from  {\sl XMM-Newton}
data  is  by far much more accurate than those determined 
from earlier observations, i.e about 60''.\\

According to the relation  between the visual extinction (A$_{\rm v}$)
and  the  hydrogen column density along the line-of-sight 
 ($\cal{N}_{\rm  H}$) reported  in
Predehl  \&  Schmitt  (\cite{Predehl95}),  we find  for  1E~1743-2843
($\cal{N}_{\rm   H}\sim$   2$\times$10$^{23}$\,cm$^{-2}$,   see
$\S$\ref{sec:spectra})  a value  of A$_{\rm  v}$ of  about  110 magnitudes.  
The source is  too absorbed to be  detected by 2MASS (Two  Micron All Sky
Survey;  http://www.ipac.caltech.edu/2mass/).  An examination  of the
maps of the NVSS, NRAO VLA Sky Survey (Condon et al. \cite{Condon98})
did  not reveal  any possible  radio counterpart  at the  position of
1E~1743.1--2843  with  flux  at  20  cm  greater  than  $\sim$50  mJy
(Cremonesi et al. \cite{Cremonesi99}).

\begin{figure}[h]
\psfig{file=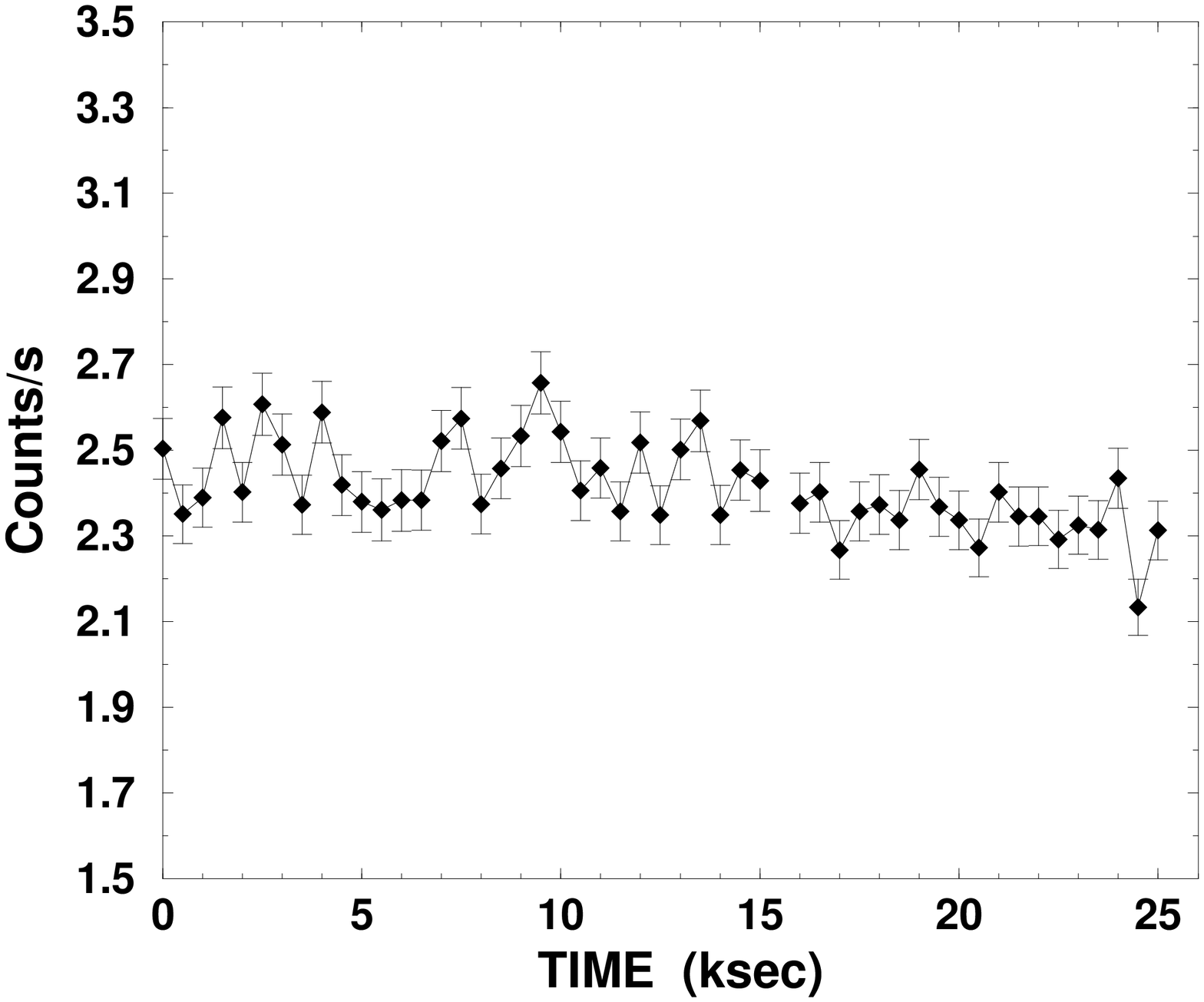,width=8cm,angle=-90}
\vspace*{-0.4cm}
\psfig{file=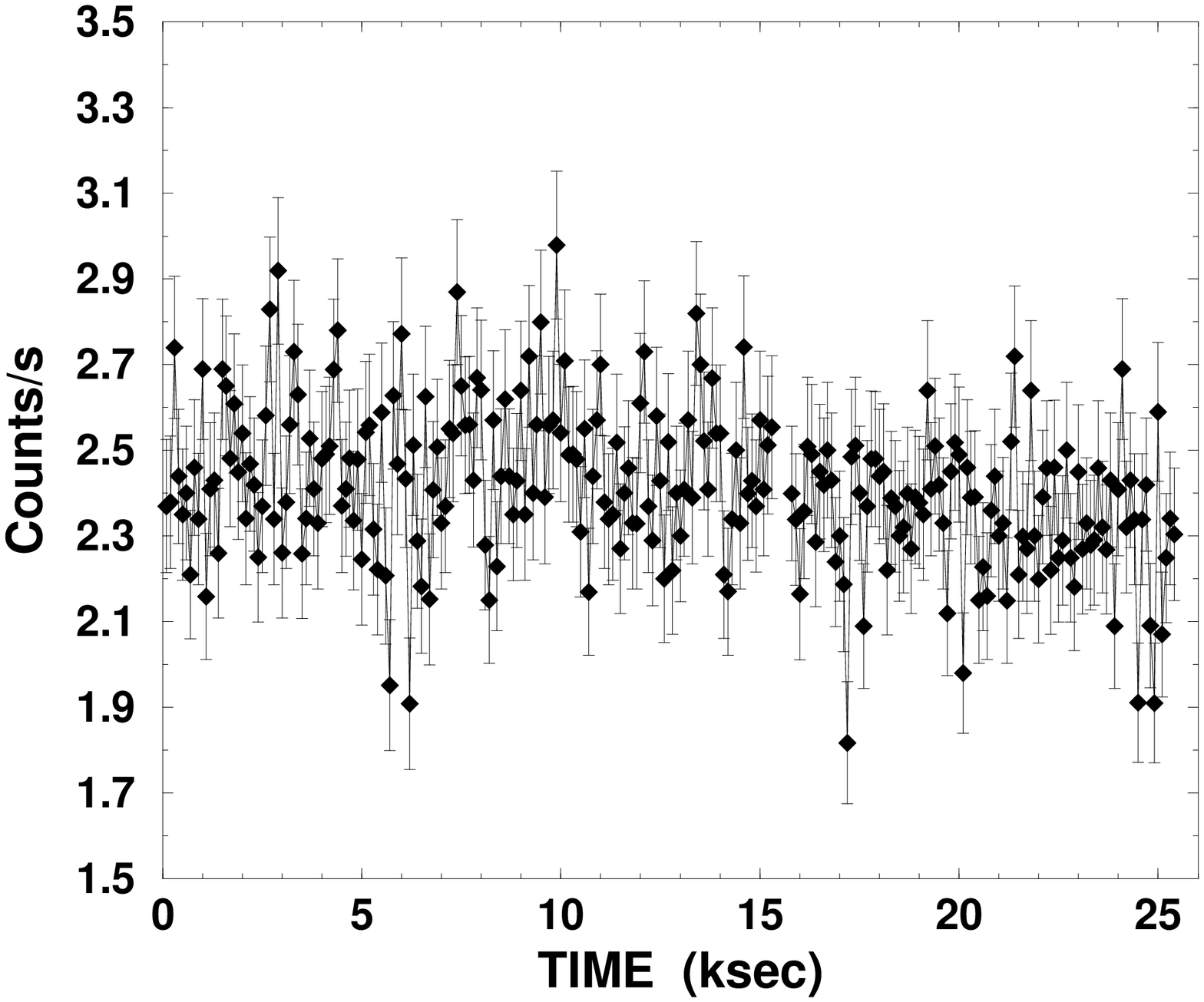,width=8cm,angle=-90}
\caption{The 2--10 keV light curve of 1E~1743-2843 measured
 with  {\sl XMM-Newton} PN detector after  background subtraction. 
{\it Top Panel}: time binning of 500\,s. 
{Bottom Panel}: time binning of 100\,s.}
\label{fig:lcmos2pn}
\end{figure}
\section{Timing analysis}\label{sec:timing}

 The 2--10\,keV background subtracted light curves of 1E~1743.1--2843
obtained  during the  {\sl  XMM-Newton} observation  are presented  in
Figure 1 for two different time binnings (100\,s and 500\,s).
 The light curve shows some variation around the mean value of
 2.42$\pm$0.10 cts\,s$^{-1}$  (Fig.\ref{fig:lcmos2pn}). 
 Fitting the light curve binned at 500\,s with a constant count rate 
gives a moderate fit with $\chi^{2}$/d.o.f=104/49. 
 In   order to determine whether or not this variability was significant,
 we produced a power spectrum using POWSPEC v1.0, between 2.4\,mHz and
2.5\,Hz.
  The resultant power spectrum is flat, and the (leahy) normalized power
density spectrum is well fit with a constant value of $1.99 \pm 0.01$
($\chi^2=25$ for 28 d.o.f.). This value is compatible with the expected
   value of 2 for a purely poissonian noise (white noise). 
 The data mode employed in the present observation allows timing
 studies only up to a frequency of 2.5\,Hz, which is quite limited,
 since many XRBs present quasi periodic variations above that value. No
 pulsations or quasi periodic oscillations (QPO) are detected in the
2.4\,mHz--2.5\,Hz
 frequency range, down to a
 relatively low level. Indeed using the relation
 $$N_{\sigma}=0.5\times\frac{S^2}{S+B}~ r^2
 \left(\frac{T}{\Delta\nu}\right)^{\frac{1}{2}},$$ with S the source net count
 rate, B the background count rate, T the exposure time, r the
 fractional amplitude,
  we can estimate a 3 sigma upper limit for a given pulsation (whose
width is equal to the frequency resolution of the power spectrum). In
our
  case, this leads to a 3$\sigma$ upper limit of $\sim 2.4\%$ for a
periodic 
pulsation and higher for any QPO (which by definition posseses a natural
width higher than 
the frequency resolution).

\indent As already pointed out by Cremonesi
et al. (\cite{Cremonesi99}), a typical Type~I X--ray burst with a peak
luminosity  close  to  the  Eddington  limit would  have  produced  in
1E~1743.1--2843 a very large count rate increase. 
 For EPIC-PN  assuming that 1E~1743.1--2843 is located
at the Galactic Center (d$\sim$8\,kpc), 
a factor of  50 increase would be observed between 
the ``quiescent state'' and the ``burst'' count rates. 
Then such  bursts would be readily seen even in 
a light curve with a time binning 
as low as 1\,s, which is very short compared to the typical duration 
 of a few tens of seconds in LMXB.
 In  addition the
non-detection of eclipses in the  X-ray light curve implies that the orbital
inclination  of the system  is smaller  than 70$^{^\circ}$  (Cowley et
al.  \cite{Cowley83}).    
We do not  see any indication of  an orbital
variation  on  long  time  scale  as suggested  by  Cremonesi  et  al.
(\cite{Cremonesi99})   but  this  is   consistent  with   our  shorter
observation duration.\\
Analysis of the Power Density Spectrum indicates that the fractional 
variability of 1E~1743.1--2843 in the frequency range 10$^{-4}$--2.5\,Hz is less than
18$\%$ rms (3 sigma upper limit). This is not a strong constraint, regarding 
the state of the source (if it is a black hole for example), as only a 
limited frequency domain has been  explored, and such a limit is compatible 
with either a soft or hard state (Nowak \cite{Nowak95}).

\section{Spectral analysis}\label{sec:spectra}

For spectral fitting, the  
data were rebinned with a minimum
of 25 counts  per bin to allow use of  the $\chi^{2}$ statistic.  {\sc
xspec}  (v11.1.0) is  used  for the  spectral  fitting.  
The
response matrice (.rmf) and  ancillary (.arf) files were computed using
the   SAS  package.    
The spectrum of 1E~1743.1--2843 was fitted between 2 and 12 keV.
The   photo-electric    absorption   cross-sections   of    Wilms   et
al. (\cite{Wilms2000})  are used throughout this paper 
 with  abundances taken from
Anders  \&  Grevesse  (\cite{Anders89}).   All errors  are
quoted  at 90$\%$  confidence. \\  \indent  We fit  the  
background-subtracted  source spectrum with various single-component  
spectral models as follows: black-body ({\sc bb}), power-law ({\sc pow}), 
 and multi-color disk black-body (MCD, {\sc diskbb}). 
 In all cases the hydrogen column density
($\cal{N}_{\rm H}$) was included as a free  parameter.  The results of this
analysis are presented in  Table~\ref{table:fits}. All of the 
simple  models noted above provided a statistical good fit to the
 observed spectrum. 

\begin{figure}
\psfig{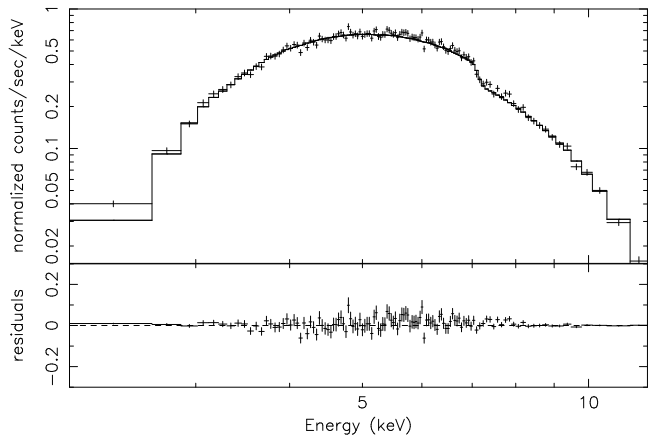}
\caption{The 2--12\,keV PN spectrum of 1E~1743.1--2843 (binning at 15$\sigma$) 
and the best-fit power-law  continuum model. 
The lower panel shows the residuals of the data to the model.}
\label{fig:spectrumXMM}
\end{figure}

\begin{table}[!ht]
\caption{The results of fitting different continuum models to the PN data 
in the 2--12\,keV energy range. 
The fluxes correspond to the 2--10\,keV band and are corrected for absorption. 
 The units are $10^{-10}$ erg\,cm$^{-2}$\,s$^{-1}$. 
}
\begin{tabular}{lcccccccc}
\hline
\hline
\noalign {\smallskip}
Models      &    ${\cal N}_{\rm H}$              &   kT (keV)          &  $\chi^{2}$/d.o.f.   & Flux \\
            &    {\tiny (10$^{23}$\,cm$^{-2}$)}  &   or  $\Gamma$ &   & {\tiny (2--10\,keV)}  \\
\noalign {\smallskip}
\hline
\noalign {\smallskip}
pow         & 2.02$\pm$0.04 &  1.83$\pm$0.05 &1111/1141         &  2.40$\pm$0.23\\
\noalign {\smallskip}
bb         & 1.31$\pm$0.03 &  1.93$\pm$0.03 &  1206/1141         &  1.52$\pm$0.02 \\
\noalign {\smallskip}
diskbb    & 1.69$\pm$0.03 &  3.4$\pm$0.1 & 1099/1141     &  1.89$\pm$0.21   \\
  \noalign {\smallskip}
\hline
\hline
\end{tabular}
\label{table:fits}
\end{table}

 For  the   single {\sc bb} and {\sc diskbb}    models,  
the  fits  are  good   but  the  inferred
temperatures  are  higher (kT$\sim$1.9\,keV and  kT$\sim$3.4\,keV, 
respectively) than  those  found  in  general for  neutron star LMXBs,  
i.e., 0.5--1.5\,keV   (e.g.,    Barret   \cite{Barret2001}).    
The parameters  found here for the bb model  
are consistent, within the error bars,  with those found
 with   {\sl BeppoSAX} in April 1998, i.e. 
 $\cal{N}_{\rm     H}$=1.3$\pm$0.1$\times$10$^{23}$\,cm$^{-2}$,
 and kT=1.8$\pm0.1$\,keV. 
 The unabsorbed  2--10\,keV  flux  found in the present 
data appears slightly lower than the one found  in April 1998, 
 i.e. 1.65$\times$10$^{-10}$\,erg\,cm$^{-2}$s$^{-1}$. \\
  Recently, absorption features associated 
to H-like and/or He-like K$_{\alpha}$ resonance lines of 
 Fe, Ca, Ne, O, and N 
 have been observed in several neutron star LMXBs 
(e.g, \object{GX 13+1}: Ueda et al. \cite{Ueda2001}; 
\object{EXO0748-676}: Cottam et al. \cite{Cottam2001};
\object{MXB1659-298}: Sidoli et al. \cite{Sidoli2001};
\object{X1624-490}: Parmar et al. \cite{Parmar2002}). 
 Such features are not statistically required 
 in the present data, with equivalent width (EW) 
 upper limits (at 90$\%$ confidence) of about 10\,eV, 1\,eV, and 15\,eV, 
respectively for \ion{Fe}{xxvi} ($\sim$ 7\,keV), \ion{Fe}{xxv} ($\sim$ 6.7\,keV), 
and \ion{Ca}{xx} ($\sim$ 4.1\,keV).
 The others lines of Ne, O, and N, below  2\,keV,
 are not accessible 
due to the very large absorption in the line-of-sight. \\

 The excellent power-law fit ($\chi_{\rm  red}^{2}$=0.975)
contrasts with the corresponding results from Cremonesi et al. (\cite{Cremonesi99}) 
 where a power-law model gave $\chi_{\rm red}^{2}$=1.49.  Moreover, fixing
the  parameters  at the  values  found  by  Cremonesi et  al.   (i.e.,
$\cal{N}_{\rm H}$=2$\times$10$^{23}$  cm$^{-2}$, and $\Gamma$=2.2), we
also obtain  a  bad   fit   for  the   power-law   model  ($\chi_{\rm
red}^{2}$=1.53).   Figure~\ref{fig:spectrumXMM} shows the  PN
spectrum  of 1E~1743.1--2843 and  the residuals of the best-fitting power-law
model to the present data.   
 This implies that 1E~1743.1--2843 could be in the present observation
  a black hole candidate  (BHC) in its {\it low/hard state} or, conceivably, an Active
Galactic  Nucleus  (AGN) observed  through  the obscuration of the
Galactic Plane.  
 The inferred photon index for 1E~1743.1--2843 is about 1.8 which is within
the   range   found  in   both   type   of   objects  (e.g.,   Wu   et
al.  \cite{Wu2001},  Malizia   et  al.  \cite{Malizia99}).   
BHC in our Galaxy are usually associated with weak   (few   mJy)  radio   
counterpart   (e.g.,   Fender  \&   Hendry \cite{Fender2000},  
Corbel et al.  \cite{Corbel2000}). For example,
the BHC \object{1E 1740.7--2942} located in the GC region, has a radio
flux at 20\,cm  of about 1.4\,mJy (Gray et  al. \cite{Gray92}), and an
unabsorbed          2--10\,keV          flux          of          about
5$\times$10$^{-10}$\,erg\,cm$^{-2}$\,s$^{-1}$        (Sidoli        et
al. \cite{Sidoli99}), which corresponds to a luminosity 
of about 5.7  $\times$  10$^{36}$ d$^{2}_{\rm 10kpc}$ erg\,s$^{-1}$.  
Similarly, Seyfert galaxies are also
  rather weak radio sources (Nagar et al. \cite{Nagar2000}). 
It follows that both the BHC and AGN hypotheses cannot be ruled out
by the radio flux limits for  1E~1743.1--2843 quoted earlier.\\
 \indent  We found that the presence of an 
  iron K$_{\alpha}$ emission line
 at 6.4\,keV (from neutral to moderatly ionized iron, i.e. $<$ \ion{Fe}{xvii}), 
 is not statistically required by our data with $\Delta\chi^{2}<$1 for one additional parameter. 
 We found an upper limit (at 90$\%$ confidence) for the EW of 12\,eV.
 This value   is   compatible   with the known properties of
LMXBs   (i.e., less than 10 to 170 eV; Asai   et al.  \cite{Asai2000}), 
 and with  extremely high luminosity  Radio-Quiet  quasars (George et al. \cite{George2000}),
 but rather weak  for a typical Seyfert  galaxy 
(EW$\sim$100-150\,eV, Nandra \& Pounds \cite{Nandra94}). 

\indent   Although the single power-law component  model
 provides a rather good fit to the PN spectrum, it is worth
investigating whether constrained 2-component models add any further information.
  We fitted a model combining  a power-law with a bb (or a MCD). 
We let all the parameters free in the fitting procedure. 
 We found a very good representation of the present data, 
 and we found a $\Delta\chi^{2}$ of about 32 for only two 
additional parameters compared to the one-component power-law 
model. The results are shown in Table~\ref{table:bbpo}.
The low value of the temperature  together with the hard spectral index 
found are compatible with a BH in a low hard state, 
as already observed for example 
 in \object{GX 339-4} (kT$\sim$0.12\,keV, Wilms et al. \cite{Wilms99}). 
 Due to the very high absorption below 2\,keV,  
we cannot  obtain a strong constraint on the normalization factor 
of the disk component (which is related to the inner radius of the disk),
 and hence no reliable constraints on the 
inner radius of a 0.1 keV accretion disk. \\

\begin{table}
\caption{The results of fitting to the PN data with two-component models, 
in the 2--12\,keV energy range.}
\begin{tabular}{ccc}
\hline
\hline
  \noalign {\smallskip}
 parameters                          &  {\sc bb+pow}                    &  {\sc diskbb+pow} \\
  \noalign {\smallskip}
\hline
  \noalign {\smallskip}
${\cal N}_{\rm H}$                 &   2.10$^{+0.07}_{-0.06}$   &  2.11$^{+0.06}_{-0.05}$   \\
{\tiny (10$^{23}$\,cm$^{-2}$)}         &                                         &                                       \\
  \noalign {\smallskip}
kT (keV)                                  &  0.15$^{+0.05}_{-0.04}$   &  0.16$^{+0.04}_{-0.05}$ \\
  \noalign {\smallskip}
$\Gamma$                             & 1.89$\pm$0.05    &   1.89$\pm$0.05 \\ 
  \noalign {\smallskip}
$\chi^{2}_{\nu}$/d.o.f.          &  1078.7/1139                    &  1078.6/1139               \\
  \noalign {\smallskip}
\hline
\hline
\end{tabular}
\label{table:bbpo}
\end{table}

\begin{figure}
\hspace*{-2mm}
\psfig{file=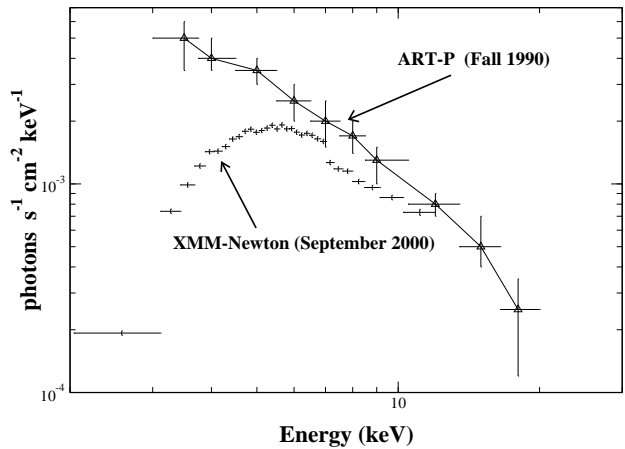,angle=-90,width=9.5cm}
\caption{Photon index spectra of 1E~1743.1--2843. 
{\it Filled circles}: present {\sl XMM-Newton} observation (power-law model). 
{\it Diamonds}: average spectrum obtained with {\sl ART-P} during Fall 1990 
(Pavlinsky et al. \cite{Pavlinsky94}).}
\label{fig:photonspectra}
\end{figure}

A  possible explanation of the differences between our spectral fits and those of
Cremonesi  et al.  (\cite{Cremonesi99}) may be  due  to a  different state  of the  object.
 We checked for possible spectral variations compared to previous 
 observations obtained from the 
X-ray coded mask telescope {\sl  ART-P} in the 4--20  keV band 
(Pavlinsky et al. \cite{Pavlinsky94}).  On the basis
of the {\sl XMM-Newton} measurements (specifically the power-law model)
 we  obtain 1.31$^{+0.13}_{-0.12}\times$10$^{-2}$\,photon\,cm$^{-2}$\,s$^{-1}$,
in the 4--20  keV band which is  smaller that the  average flux determined 
using  {\sl ART-P} during Fall 1990 
(1.77$\pm$0.08 $\times$10$^{-2}$\,photon\,cm$^{-2}$\,s$^{-1}$; Table~1
in Pavlinsky et al. \cite{Pavlinsky94}).  
In fact 1E~1743.1--2843 is known to be variable in hard X-rays from
ART-P observations carried out from Spring 1990 to Winter 1992.
Figure~\ref{fig:photonspectra}  compares the photon spectrum 
inferred from the {\sl XMM-Newton} observation
with the average photon  spectrum measured by {\sl
ART-P}  during Fall  1990.  
In the more recent observation, the intensity
is significantly lower at energies  below 
6\,keV. For example the flux at 3.5\,keV is about
 7 times lower than during fall 1990.
The difference could be due to a change 
in absorption between the two observations. 
 Above 6\,keV the spectral slope appears flatter.  
This is at least weak evidence for the fact that the source 
changes of state from time to time. \\

The observation of 1E~1743.1--2843 at higher energies (E$>$ 30  keV)  
would  certainly help  to determine the  nature  of  the  object, 
  indeed up to 12\,keV a very absorbed black-body model 
with kT about 1.8\,keV and a very absorbed power-law model 
have similar shapes.
 Unfortunately  no  significant detection  above 30\,keV
 of  1E~1743.1--2843 has   ever  been   obtained  (see   e.g.   Goldwurm  et
al. 1994). This has previously been justified in terms of the source 
having a soft thermal spectrum; however the {\sl XMM-Newton} 
observation indicates that the source can enter states where the spectral
form is relatively hard.

\section{Summary and discussion}

We report  on {\sl XMM-Newton} observation of the  bright 
X-ray source 1E~1743-2843. 
  During  the observation  the  source  flux remained relatively 
steady at a level corresponding to a  2--10\,keV   
luminosity         of      about       2.7      $\times$
10$^{36}$~d$_{10\rm{kpc}}^{2}$\,erg\,s$^{-1}$ (assuming a power-law continuum).  
 As in  previous X-ray
observations, there was no evidence for either X-ray  bursts, strong
chaotic variability  or  significant  pulsations.\\ 

\indent The {\sl XMM-Newton}  spectrum of 1E~1743.1--2843 
can be well fitted with simple
 very absorbed (${\cal N}_{\rm H}$=13--20 $\times$ 10$^{22}$\,cm$^{-2}$), 
featureless, one-component emission models, 
such as power-law continuum. 
We found that the present data are compatible 
with  a black  hole candidate in  a 
low/hard state. The fact that the measured spectrum  is slightly  harder 
than measured earlier by  {\sl BeppoSAX} and {\sl ART-P} observations
is a tentative indication to a change of state but our  spectral coverage 
is too limited  to draw a firm conclusion.  
 The hypothesis of  a neutron star LMXB,  
as proposed earlier by Cremonesi et al. (\cite{Cremonesi99}), 
is not ruled out by our spectral analysis.  
It could be also an extragalactic source  seen through the Galactic Plane.  
Unfortunately,  the data mode employed in the  present observation 
is not suitable for advanced  timing analysis,  in particular  
we are unable to investigate whether the
source exhibits millisecond X-ray pulsations, which may be detectable from
neutron star LMXBs. \\

It is  noteworthy that the X-ray source  1E~1743.1--2843 is located within 
$20^{\prime}$ of the Galactic Center and, in projection, lies 
on the periphery of SNR \object{G0.33+0.04} where the SNR emission 
is the brightest at 90\,cm (Kassim \&  Frail \cite{Kassim96}).  
Also 1E~1743.1--2843  lies very  close,  again in projection,
to  a  giant  molecular cloud  core
\object{GCM+0.25+0.01}  ($\alpha_{\rm   J2000}$=17h\,46m\,10.1s,
$\delta_{\rm J2000}$=--28$^{\circ}$\,42$^{\prime}$\,48.4$^{\prime\prime}$),
which appears to  be located at  the GC region,  and to contain embedded  
low-mass star formation (Lis et al.  \cite{Lis94}).  
If 1E~1743.1--2843 is located in the  GC region, and  behind this  cloud, 
this  could explain  the high absorption  of its soft  X-ray flux.   
In such circumstances we might expect to  observe significant  
6.4\,keV  fluorescent line  emission in  this  
cloud due  to  the  high  X-ray illumination  from 1E~1743.1--2843.  
Indeed, Lis  et al. (\cite{Lis94}) have already remarked that this  
cloud might be a further  example of an  X-ray irradiated reflection
nebula.   If no significant Fe\,K$_{\alpha}$ line at 6.4\,keV 
(neutral or moderatly ionized iron) is detected, 
 then one can  infer that 1E~1743.1--2843 is far from the Galactic Center region, 
i.e. at a distance greater than 8 kpc, and has a X-ray luminosity 
  higher than 10$^{36}$\,erg\,s$^{-1}$.
 A mosaic with a higher S/N of the Galactic Center region may answer 
this point (Decourchelle et al. 2003, in preparation).  
 According to  the formula  of Sunyaev  \& Churazov
(\cite{Sunyaev98}) and assuming a minimum projected distance,  
 we   obtained    a   line    flux    of   about  
10$^{-5}$\,erg\,s$^{-1}$, i.e. approximatively  6 times lower than the
very  bright  line emission observed from the giant  molecular  cloud  Sgr\,B2
(5.6$\times$10$^{-5}$\,erg\,s$^{-1}$,            Murakami           et
al.  \cite{Murakami2001}).  Then 1E~1743.1--2843  could also be a contributor 
in a smaller part as shown above to the 6.4\,keV, iron line observed  in Sgr B2 (1E
1743.1--2843   which lies  63\,pc away, in terms of the minimum projected
distance,   Murakami  et al. \cite{Murakami2000}).  
Long-term monitoring of
1E~1743.1--2843 coupled with a more detailed investigation into
its interaction with GC molecular clouds (if any) could therefore be
useful for the development of a more complete view of GC activity.\\

The greatly improved positional constraints from {\sl XMM-Newton}   
 should, in the future, help in the search for a possible counterpart  which, 
in term, would  contribute greatly  to our understanding of its true nature.
 Observations  at higher energies  are needed  to determine  the source
spectral state, but up  to now the location of the source in the crowded 
Galactic Center region and the limited sensitivity and spatial
resolution of the available instrumentation have hampered such an 
investigation.  Observations with the IBIS
instrument onboard the {\sl  INTEGRAL} mission, will bring information
about the hard  X-ray (i.e.  $>$\,30\,keV) and possible gamma-ray  emission 
of 1E~1743.1--2843 without any spatial  confusion, which should help in
the determination of its  nature.   In particular,  significant 
 hard X-ray and gamma-ray emissions
  are expected from BHC in low/hard states.

\begin{acknowledgements}
This work  is based on  observations obtained with {\sl XMM-Newton},  an ESA
science mission with instruments  and contributions directly funded by
ESA Member  States and the  USA (NASA).  
We would like to thank the anonymous referee for a careful reading 
of the manuscript. 
D. P.  and J. R. acknowledges financial
support from a postdoctoral  fellowship from the French Spatial Agency
(CNES). 
\end{acknowledgements}

\end{document}